\documentclass[twocolumn,10pt]{IEEEtran}

\usepackage[absolute]{textpos}
\usepackage{everyshi}

\usepackage{indentfirst}
\usepackage{amsmath}
\usepackage{amsfonts}
\usepackage{amssymb}
\usepackage{cite}
\usepackage{times}
\usepackage{booktabs}
\usepackage[dvips]{graphicx}
\usepackage{hhline}
\usepackage{color}
\usepackage{multirow}
\usepackage{makecell}
\usepackage[table]{xcolor}
\usepackage{graphicx}
\usepackage{epstopdf}
\usepackage{algpseudocode}
\usepackage{algorithm}
\usepackage{adjustbox}
\usepackage{caption, subcaption}

\usepackage{url}

\usepackage[shortlabels]{enumitem}

\usepackage{color,soul}

\newcommand{\bi}{\begin{itemize}}
\newcommand{\ei}{\end{itemize}}

\newcommand{\be}{\begin{IEEEeqnarray}}
\newcommand{\ee}{\end{IEEEeqnarray}}

\newcommand{\commentLB}[1]{}
\usepackage[acronyms,nonumberlist,nopostdot,nomain,nogroupskip]{glossaries}
\newacronym{eeg}{EEG}{electroencephalography}
\newacronym{eog}{EOG}{electroencephalogram}

\newacronym{car}{CAR}{Common Average Reference}

\newacronym{ica}{ICA}{Independent Component Analysis}

\newacronym{bci}{BCI}{Brain-Computer Interface}

\newacronym{mrcp}{MRCPs}{Movement Related Cortical Potentials}

\newacronym{ml}{ML}{machine learning}
\newacronym{cv}{CV}{cross-validation}

\newacronym{lda}{LDA}{Linear Discriminant Analysis}
\newacronym{slda}{sLDA}{shrinked linear discriminant analysis}

\newacronym{rf}{RF}{random forest}
\newacronym{cnn}{CNN}{convolutional neural network}

\newacronym{rbf}{rbf}{radial basis function}
\newacronym{svm}{SVM}{Support Vector Machine}
\newacronym{fft}{FFT}{Fast Fourier Transform}
\newacronym{nn}{NN}{neural network}

\newacronym{elu}{eLu}{exponential linear unit}

\begin{document}

 \title{Deep learning-based classification of fine hand movements from low frequency EEG}

 \author{
   Giulia~Bressan$^{1}$, Selina~C.~Wriessnegger$^{2}$, Giulia~Cisotto$^{1,3,4}$,~\IEEEmembership{Member,~IEEE} \\
   
   \small{
   $^{1}$Dept. of Information Engineering, University of Padova, Padova, Italy \\
   $^{2}$ Institute of Neural Engineering, Graz University of Technology, Graz, Austria \\
   $^{3}$ National Center of Neurology and Psychiatry, Tokyo, Japan \\
   $^{4}$ CNIT, the National, Inter-University Consortium for Telecommunications \\
   email: $\{$giulia.bressan.9@studenti.unipd.it, giulia.cisotto.1@unipd.it, s.wriessnegger@tugraz.at$\}$
   }
   }

 \maketitle
 \thispagestyle{empty} \pagestyle{empty}

 \begin{abstract}
The classification of different fine hand movements from EEG signals represents a relevant research challenge, e.g., in brain-computer interface applications for motor rehabilitation.
 Here, we analyzed two different datasets where fine hand movements (touch, grasp, palmar and lateral grasp) were performed in a self-paced modality.
 We trained and tested a newly proposed \gls{cnn}, and we compared its classification performance into respect to two well-established machine learning models, namely, a shrinked-LDA and a Random Forest.
 Compared to previous literature, we took advantage of the knowledge of the neuroscience field, and we trained our \gls{cnn} model on the so-called \gls{mrcp}s. They are EEG amplitude modulations at low frequencies, i.e., $(0.3, 3)$ Hz, that have been proved to encode several properties of the movements, e.g., type of grasp, force level and speed.
 We showed that \gls{cnn} achieved good performance in both datasets and they were similar or superior to the baseline models. Also, compared to the baseline, our \gls{cnn} requires a lighter and faster pre-processing procedure, paving the way for its possible use in an online modality, e.g., for many brain-computer interface applications.
\end{abstract}

%

\begin{textblock*}{17cm}(1.7cm, 0.5cm)
	\noindent\scriptsize This work has been submitted to the IEEE for possible publication. Copyright may be transferred without notice, after which this version may no longer be accessible.\\
	\textbf{Copyright Notice}: \textcopyright 2020 IEEE. Personal use of this material is permitted. Permission from IEEE must be obtained for all other uses, in any current or future media, including reprinting/republishing this material for advertising or promotional purposes, creating new collective works, for resale or redistribution to servers or lists, or reuse of any copyrighted component of this work in other works.
\end{textblock*} 


 \glsresetall



 \section{Introduction}\label{sec:intro}
 The recognition and classification of different fine hand movements from \gls{eeg} signals represents an interesting and challenging research question.
Several \gls{bci} systems for motor rehabilitation \cite{Cisotto2013, Silvoni2013, Cisotto2014} and other basic neuroscience studies, such as the investigation of the neural mechanisms underlying the writing and the music performance \cite{Furuya2016, CisottoSfN2017}, strongly rely on the ability to precisely and effectively distinguish different fine hand movements.
\gls{mrcp} are amplitude modulations of the time-domain EEG signal, that occur in the $(0.5, 4)$ Hz frequency band \cite{shibasaki2006bereitschaftspotential}. \gls{mrcp} can be detected during motor execution and imagery, or in an attempted movement, and they reflect the cortical processes involved in the planning and execution of a movement.
Previous literature \cite{shibasaki2006bereitschaftspotential} reports that the components of the \gls{mrcp} can be influenced by several factors, such as the preparatory state (self-paced or cue-based), the level of intention, the type of movement, the praxis and the previous experience of the same movement, besides the presence of any pathology of the brain structures.
Nevertheless, it has been found \cite{pereira2018eeg} that \gls{mrcp} can also encode several properties of the movements, such as the type of grasp, the force level and the speed of the task.
For this reason, for example, \gls{mrcp} are considered valid signals to be used for \gls{bci} control \cite{pereira2018eeg}.
Here, we fused this knowledge from the neuroscience field, with the potentiality of deep learning, to improve the performance of the classification of touch, grasp, palmar and lateral grasp movements.
Previous literature has already investigated the classification of different fine hand movements, including touch and different kinds of grasp.
The majority of the studies employed \gls{slda}, which is a well-established approach for \gls{eeg} classification, for its low complexity and good performance even with a limited amount of trials.
However, \gls{lda} and its regularized version, \gls{slda}, are linear classifiers which might score poorly in case of complex non-linear \gls{eeg} data \cite{lotte2007review}.
The aim of this work was to evaluate the performance of a newly proposed \gls{cnn} model, in comparison with two standard machine learning algorithms, namely \gls{slda} and \gls{rf}, in the classification of $3$ different classes of movement, using two datasets.
The paper is organized as follows.
In Section~\ref{sec:related-works} we present the most relevant previous studies related to our work; in Section~\ref{sec:methods}, we describe the experimental protocol, the common steps of pre-processing for all models, our proposed \gls{cnn} model, and we briefly review the two baseline models chosen as a comparison for performance evaluations. In section~\ref{sec:results} we report and discuss all results, both from the qualitative analysis of the \gls{mrcp} and of the classification of different movements. Finally, Section~\ref{sec:conclusions} concludes the paper, also mentioning the possible impact of this work for other studies.

 \section{Related works} \label{sec:related-works}
 The possibility of decoding touch and grasp actions from low-frequency EEG signals has been shown in other studies \cite{ofner2017upper, schwarz2017decoding, schwarz2019unimanual}.
In \cite{ofner2017upper}, Ofner \emph{et al.} classified single upper limb movements with a binary classification approach, recording six different types of movements, both executed and imagined, and rest trials. For the executed movements, in the \textit{movements} \textit{versus} \textit{rest} binary classification the average accuracy reached the value of $87\%$, while for the \textit{movements} \textit{versus} \textit{movements} the average accuracy dropped down to $55\%$. For the imagined movements, an accuracy of as less as $27\%$ and $73\%$ was obtained for \textit{movements versus movements} and \textit{rest versus movements} classification, respectively.
In \cite{schwarz2017decoding} palmar, lateral and pincer grasps were recorded and classified, in a cue-based paradigm. A $4$-class \gls{slda} was used to classify the three movements and the rest data, obtaining a peak accuracy of $65.9\%$. Moreover, a binary classifier was trained in the same study, for each binary combination of classes. The palmar versus lateral grasp classification obtained a peak accuracy of $73.5\%$.
In \cite{schwarz2019unimanual}, both unimanual and bimanual reach and grasp actions were classified with \gls{slda}. Binary combinations of the different movements were also classified separately, leading to average accuracies for the movement classes between $66\%$ and $70\%$. The highest accuracies were obtained with the \textit{rest} class versus the \textit{movement} ones, with performance between $74\%$ to $90\%$.
Recently, new approaches have been rising. Deep learning showed promising results in many different fields of application and has been successfully applied also in the \gls{bci} field \cite{lotte2007review}.
%

 \section{Methods}\label{sec:methods}
 In this section, we first present the experimental protocol used to acquire the two datasets. Second, we describe the common pre-processing pipeline that is used by \gls{cnn} and the baseline models. Then, we introduce our \gls{cnn}-based model and the baseline models used for the performance comparison.
Finally, we explain the cross-validation procedure and the metric we used as for the evaluation of the performance.

\subsection{Experimental protocol}\label{sec:expprotocol}

At the very beginning of the experimental protocol, the participants' handedness was tested with the well-known hand dominance test of \cite{steingruber1971hand}.
Then, they were asked to seat on a comfortable chair in a noise and electromagnetic shielded room.
The brain activity was acquired via \gls{eeg} by means of $4$ g.USBamp amplifiers (g.tec
medical engineering GmbH, Austria) and a $64$~gel-based channel \gls{eeg} cap (g.GAMMAsys/g.LADYbird, g.tec
medical engineering GmbH, Austria). Incidentally, $58$ electrodes recorded the brain activity, while $6$ of them were used to record the \gls{eog}. The \gls{eeg} electrodes' locations were defined by a well-established modified version of the International $10-20$ System~\cite{oostenveld2001five}.
All data were recorded using a $256$~Hz sampling frequency.
In the resting position, the participants' right arm was placed, relaxed, upon a pressure button on a table in front of them. They were also recommended to avoid unnecessary body or eye movements, and to fix their gaze at a fixed point, for a few seconds, at the beginning of each repetition of the movement.
All movements were self-initiated to ensure a more natural application scenario.
Additionally, at beginning, middle and end of the experiment, $3$~min rest is repeated $3$ times.

\subsubsection*{Experiment 1 - Touch and Grasp}\label{sec:exp1}
In the first experiment, $11$ healthy volunteers (age 20-38 years old, $11$~M) were included. The hand dominance test resulted in $9$ right-handed participants, $1$ left-handed and $1$ undefined.
During the experiment, two glasses were on the table at the participant's reaching distance. They were equipped with a pressure sensor, each, in order to precisely detect the grasping onset.
The participants were instructed either to grasp the first glass or to touch the second glass for a minimum time of $4$~s. Thus, the total duration of each repetition was longer than $5$~s.
Four sessions of $20$ repetitions of the same movement, i.e., grasping and touching, were included in the protocol. Thus, $80$ touching and $80$ grasping movements were performed by each participant at the end of the experiment.
After each session, the participants could take a break and the glasses were switched. The same number of repetitions was performed in both glass' positions.
On the computer screen in front of them, they could see the remaining number of trials to perform.

\subsubsection*{Experiment 2 - Palmar and Lateral}\label{sec:exp2}
In the second experiment, $15$ right handed participants were involved. During the experiment, two jars were on the table at the participant's reaching distance. The first one was empty, while the second had a spoon stuck in it.
The participants were instructed either to reach-and-grasp the first jar or the second for a minimum time of $2$~s. Thus, the total duration of each repetition was longer than $5$~s. They freely decided which movement to perform.
To interact with the empty jar, they had to perform a palmar grasp, while for the jar with the spoon, they exploited a lateral grasp.
Four sessions of $20$ repetitions of the same movement, i.e., palmar or lateral grasp, were included in the protocol. 
After each session, the participants could take a break and the objects were switched. The same number of repetitions was performed in both objects' positions.

\subsection{Pre-processing}\label{sec:preproc}
We adopted the same pre-processing pipeline for both the \gls{eeg} datasets used in this study. The pipeline is a well-established algorithm, previously implemented in~\cite{schwarz2017decoding, schwarz2019unimanual}.
The full data processing was implemented in Matlab 2020a~\cite{MATLAB:2020a}. 
First, every \gls{eeg} signal was band-pass filtered between $0.01$~Hz and $100$~Hz (Chebyshev filter, order $8$).
Second, a notch filter was applied to suppress the power line noise at $50$~Hz.
Additionally, \gls{ica} could be applied at this point to identify and remove the artifacts due to eye movements, as in~\cite{makeig1996independent}.
Third, a narrower band-pass filter (Butterworth filter, order $4$) was applied to extract the signal low-frequency component in the band $(0.3, 3)$~Hz.
All filters were implemented using the Matlab function \emph{filtfilt} in order to compensate for the delay introduced by them.
The full dataset, i.e., including all \gls{eeg} signals, was transformed using the \gls{car} filter~\cite{ludwig2009using}, a spatial filter used to enhance the signal component due to the brain region under each individual \gls{eeg} sensor (i.e., discarding components that are spread all around the scalp).
Finally, every signal was downsampled to $16$~Hz (using the Matlab function \emph{resample}). 

During the experimental sessions, a pressure sensor (either on the table or on the object to interact with, see Section~\ref{sec:expprotocol}) was exploited to identify the time instants when the individual initiated the movement, i.e., the movement onset.
Therefore, proper segmentation of the continuous pre-processed \gls{eeg} signals was ensured.
Each segment (or \emph{trial}) was defined as the signal's period of time from $-2$~s to $+3$~s around each movement onset (i.e., time $0$).
Not only movement-related trials, but also $5$~s \emph{rest trials} have been obtained from the datasets: they were extracted from the $3$~min rest periods (see Section~\ref{sec:expprotocol}).

In order to include only clean data in the datasets to analyse, we applied a well-established outlier rejection algorithm~\cite{schwarz2015co, statthaler2017cybathlon, faller2012autocalibration}. A single trial was kept in the dataset if it simultaneously met the following conditions: (1) its absolute amplitude does not exceed $125 \mu V$, (2) and its kurtosis does not exceed its standard deviation by $4$ times.

Finally, we obtained two different $3$~class datasets: \emph{dataset 1} includes clean data from \emph{Experiment 1}, while \emph{dataset 2} includes those from \emph{Experiment 2}.
Both dataset can be described as follows:
\begin {equation}
\mathbf{X}^{(i)} =   \left[
                    \begin{matrix}
                         x^{(i)}_1(1)  & x^{(i)}_1(2) & \cdots & x^{(i)}_{1}(N) \\
                         x^{(i)}_2(1)  & \ddots    & \cdots & \vdots    \\
                         \vdots        & \vdots    & \ddots & \vdots    \\
                         x^{(i)}_{58}(1)  & \cdots    & \cdots & x^{(i)}_{58}(N) \\
                    \end{matrix}
                \right]
                \label{eq_correlation}
\end{equation}

where $i$ is the total trial number (including all classes of movements), and $N$ the number of time samples available.
To note, $N$ varies depending on the learning model used to analyse the data (see Sections~\ref{subsec:cnn} and~\ref{subsec:baseline}).
$\mathbf{X}^{(i)}$ can be interpreted as an \gls{eeg} $2$D image.

Moreover, the class of movements can be either \emph{touch}, \emph{grasp}, \emph{palmar}, \emph{lateral} or \emph{rest}. \emph{Dataset 1} includes \emph{touch}, \emph{grasp} and \emph{rest} classes, while \emph{dataset 2} includes \emph{palmar}, \emph{lateral} and \emph{rest} classes.

\subsection{Classification with \gls{cnn}}\label{subsec:cnn}
\begin{figure}[h!]
\centering
\includegraphics[width=1\columnwidth]{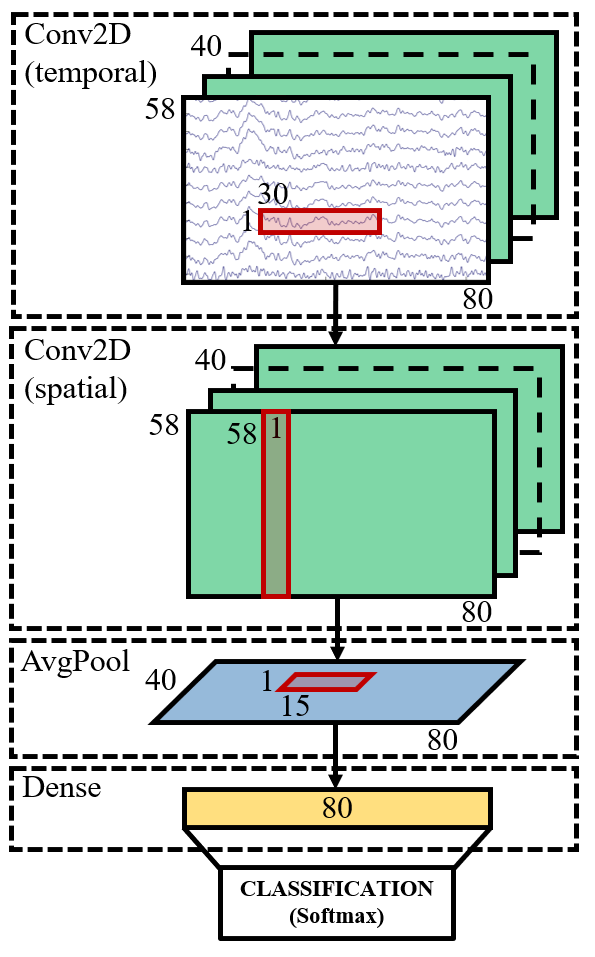}
\caption{Schematic representation of the proposed \gls{cnn} model architecture.}
\label{fig:CNNimage}
\end{figure}

The \gls{cnn} is a particular type of neural network that implements, in at least one of its layers, a convolutional operation \cite{goodfellow2016deep}.
%
%
In this study, the architecture of the \gls{cnn} was adapted from \cite{dose2018end, lee2020motor}.
As depicted in Fig.~\ref{fig:CNNimage}, it consisted of $7$ layers.
The first two were convolutional layers: the first one performed a temporal filtering (i.e., convolution along the time axis), while the second one a spatial filtering (i.e., convolution along the channel axis).
Each convolutional layer was followed by a batch normalization layer and an \gls{elu} activation function.
Then, an average pooling layer, which flattened the input to a single dimension, and two fully connected layers were stacked on the top of the convolutional ones. Finally, a softmax activation function returned the probability of each sample to belong to each class.
To note, since the kernel size at the output of the second convolutional layer was equal to the number of channels, this filter reduced the channel dimension to one.
%
The input to this \gls{cnn} was given by the EEG 2D images $\mathbf{X}^{(i)}$, for every available trial $i$, as computed in eq.~(\ref{eq_correlation}), which resulted in a three dimension tensor.
To implement such architecture, several parameters had to be decided: specifically, the kernel size and the depth of the convolutional layers, and the size of the pooling and dense layers.
Given each participant, we used a grid-search procedure to optimize such parameters over a-priori selected ranges. Then, the optimal combination of parameter values was given by a majority vote strategy across all participants.
As a result, the kernel size of layer $1$ (i.e., the first convolutional layer) was equal to $30$, while for layer $3$ (i.e., the second convolutional layer) corresponded to the number of channels, i.e., $58$. Moreover, for both of them the optimal depth was found to be $40$ filters.
The kernel size of the average pooling layer was equal to $15$, the first fully connected layer had $80$ neurons, while the second fully connected layer had $3$ neurons, corresponding to the number of classes.
The same \gls{cnn} architecture was used for both datasets, while one \gls{cnn} model was trained for each participant.

%
%
%
%

\subsection{Classification with baseline models}\label{subsec:baseline}
Two state-of-the-art machine learning models were used as a comparison for our proposed \gls{cnn}: an \gls{slda} and a \gls{rf}. They both have the advantages to be simple in their implementation, computational light burden, and they showed good performance in \gls{eeg} classification during hand movements, gesture recognition and \gls{bci} experiments.

The \gls{lda} is a supervised multi-class classification technique which aims at estimating the parameters of the linear multivariate model of the input data, via parametric density estimation procedure \cite{lotte2007review}. 
%
%
%
%
Here, the input to the \gls{slda} is the vector $\mathbf{x}$ obtained by reshaping matrix $\mathbf{X}$ as follows:
\begin{equation}
\mathbf{x}^{(i)} = [x^{(i)}_1(1),x^{(i)}_2(1),...,x^{(i)}_{58}(1), x^{(i)}_1(2),x^{(i)}_2(2),...,x^{(i)}_{58}(N)],
\label{eq:x_sLDA}
\end{equation}
where $i$ is the trial number, and $N$ the number of time samples available in the sliding window.
The shrinked \gls{lda} version, i.e., the \gls{slda}, introduces a regularization strategy, especially useful with high dimensional feature spaces, when only a few data points are available.
For the regularization, we considered the pooled covariance matrix, computed from the $3$ classes, and we optimized the regularization parameter as in \cite{bartz2014covariance}.
%
%
A common approach to obtain the optimal \gls{slda} model with time series, i.e., as in the \gls{eeg} case, is to train several \gls{slda} models, each one based on a different subset of the training set (e.g., given by a different observation window), and to select the one which yields the best training performance.
Thus, here, for each single trial $i$, a sliding window is used to scan the entire \gls{eeg} segment from $-2$~s to $3$~s. Then, an \gls{slda} model was obtained for each, every $2$, time instant (i.e., one every $125$~ms).
For each participant, the time instant where the \gls{slda} model resulted in the best classification performance was taken as the trained model. Moreover, three different window lengths were tested for each participant, specifically $\{0.6, 0.8, 1\}$~s, and the same model training was repeated for every length value.
%
%
%


The \gls{rf} is a classifier that works as an ensemble of individual decision tree algorithms to reduce the risk of overfitting and, thus, to enhance the classification performance.
Each tree is obtained by independently bootstrapping the samples from the input dataset, resulting in uncorrelated models whose predictions are more accurate than the ones we would obtain from a single one \cite{shalev2014understanding}.
Then, a random set of predictors is used at each split to grow the tree \cite{breiman2001random}.
To compute the predictions, a majority vote across the predictions of the individual decision trees is used.
In this study, the vector in eq.~(\ref{eq:x_sLDA}) was also used as the input to the \gls{rf}. The number of trees was empirically set to $50$, found as the best trade-off between the classification accuracy and the computational complexity.

\subsection{Cross-validation and performance evaluation}\label{sec:eval}
The performance of the classifiers were evaluated by means of the accuracy, computed as follows: $$\mbox{accuracy} = \frac{\mbox{correctly classified instances}}{\mbox{total number of instances to classify}}.$$
The chance level was computed for each model and each participant by means of the \emph{Adjusted Wald Interval} \cite{muller2008better}, with $\alpha$ set to $0.05$.
For both datasets, we split each of them into a training set ($75\%$) and a validation set ($25\%$).
During training, a $10$ times repeated $5$-fold cross validation procedure was adopted to ensure the robustness of the trained model.
The validation set was used for testing the performance of the trained models on unseen data.
All splits led to representative subsets of the dataset, in order to have balanced classes for an unbiased classification.

\section{Results and Discussion}\label{sec:results}
In this section, we describe both the quality of our dataset after pre-processing and the results of the classification using the \gls{cnn} model designed in Section \ref{subsec:cnn}, including the comparison with \gls{slda} and \gls{rf}.
\subsection{Pre-processing, feature extraction and \gls{mrcp}}
\begin{figure}[h!]
\centering
\includegraphics[width=1\columnwidth]{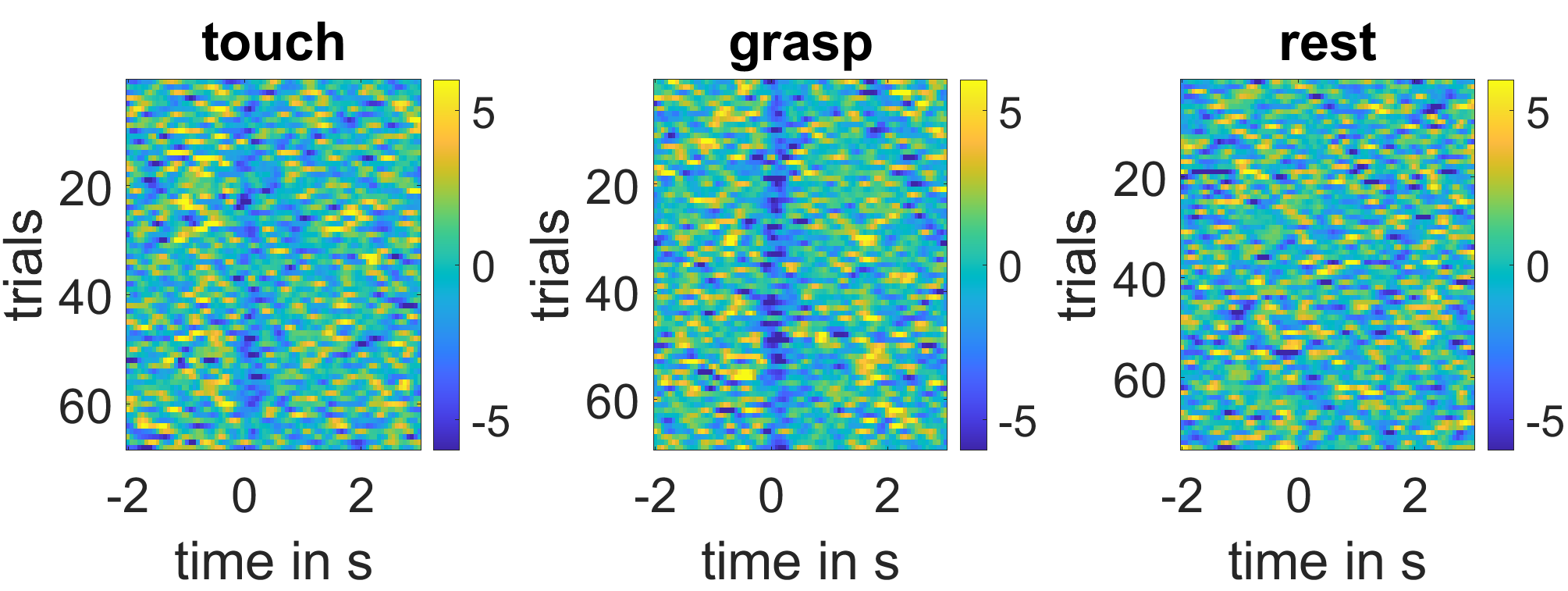}\\(a)
\includegraphics[width=1\columnwidth]{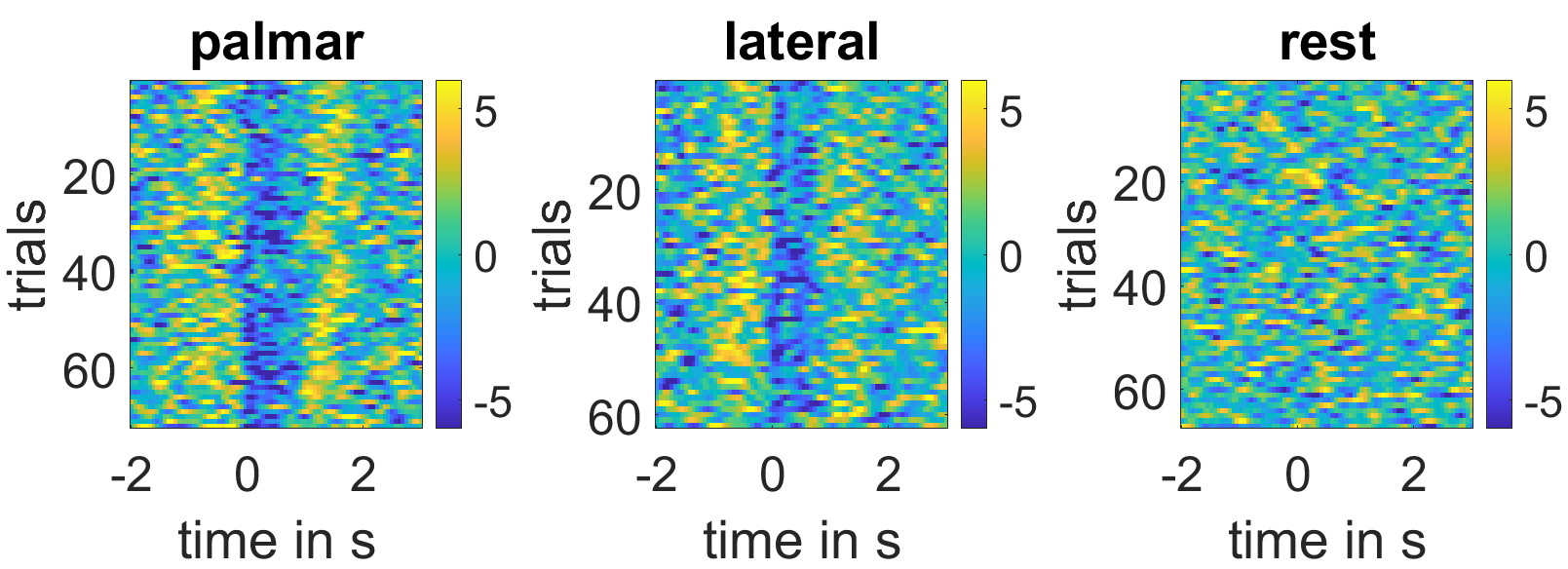}\\(b)
\caption{\gls{eeg} segments after synchronization at the movement onset. (a) Dataset 1, representative participant \emph{Subject S000} and channel $C1$. (b) Dataset , representative participant \emph{Subject G04} and channel $C1$.}
\label{fig:mrcp}
\end{figure}

As a result of the pre-processing (see Section~\ref{sec:preproc}), $3$ out of $11$ participants (namely, $S002$, $S003$, $S005$) were rejected from the \emph{dataset 1} from further analysis, for the massive presence of artifacts in their \gls{eeg} recordings.
Then, the high quality of the clean \gls{eeg} data after pre-processing is shown in Fig.\ref{fig:mrcp}. It reports the subset of \gls{eeg} segments, after synchronization at the movement onset, for different movement classes and for rest periods, in both datasets.
%
%
In Fig.\ref{fig:mrcp}, we can notice that, in case of any movement, negative values are seen around time zero, i.e. the movement onset, which represent the negative peak of the \gls{mrcp}. Moreover, all panels show good repeatability across movement repetitions (i.e., segments). On the opposite, as expected, across the rest segments we cannot notice any clear pattern.
We also observed (results not reported for space constraints) that a difference in the \gls{mrcp} peak amplitude was especially noticeable at the \gls{eeg} electrodes located in the contralateral side of the movement and that this spatial pattern is consistent across several participants, in line with other literature \cite{rice2007line}.
However, it is also clear that \emph{Dataset 1} is more affected by noise compared to \emph{Dataset 2}, so that e.g., the touch-related \gls{eeg} data could show a less pronounced negative peak of the \gls{mrcp} (as seen in Fig.\ref{fig:mrcp}a).
We also observed that this behaviour is consistent across most of the channels, with no specific spatial pattern (results not reported for space constraints).
%
%
%
%
%
%

\begin{table*}[htbp]
\tiny
\centering
\caption{Comparison of classification performance (in terms of accuracy) in validation from \emph{Dataset 1}. Different window lenghts were tested for \gls{slda} and \gls{rf}.}
\resizebox{0.9\textwidth}{!}{
\begin{tabular}{cccccccc}
\toprule
\textbf{Subject} & \textbf{sLDA (0,6s)} & \textbf{sLDA (0,8s)} & \textbf{sLDA (1s)} & \textbf{RF (0,6s)} & \textbf{RF (0,8s)} & \textbf{RF (1s)} & \textbf{CNN}  \\
\midrule
S000             & 0,60              & \textbf{0,64}              & 0,64            & 0,58            & 0,58            & 0,58          & 0,62 \\
S001             & 0,50              & 0,52              & 0,60            & 0,54            & 0,56            & \textbf{0,69} & 0,66          \\
S004             & \textbf{0,75}     & 0,73              & 0,75            & 0,69            & 0,73            & 0,69          & 0,74          \\
S006             & 0,76              & 0,72              & 0,70            & \textbf{0,86}   & 0,74            & 0,82          & 0,84          \\
S007             & 0,76              & \textbf{0,80}     & 0,76            & 0,78            & 0,73            & 0,80          & 0,68          \\
S008             & 0,78              & 0,78              & 0,83            & 0,88            & 0,80            & \textbf{0,93} & 0,85          \\
S009             & 0,62              & \textbf{0,67}     & 0,50            & 0,62            & 0,58            & 0,62          & 0,61          \\
S010             & 0,52              & 0,54              & 0,57   & 0,52            & 0,50            & 0,46          & \textbf{0,59}          \\
\midrule
\textbf{MEAN}    & 0,66              & 0,68              & 0,67            & 0,68            & 0,65            & \textbf{0,70} & \textbf{0,70} \\
\textbf{STD}    & 0,11              & 0,10              & 0,11            & 0,14            & 0,11             & 0,15          & 0,10         \\
\bottomrule
\end{tabular}}
\label{tab:AccComp1}
\end{table*}

\begin{table*}[htbp]
\tiny
\centering
\caption{Comparison of classification performance (in terms of accuracy) in validation from \emph{Dataset 2}. Different window lenghts were tested for \gls{slda} and \gls{rf}.}
\resizebox{0.9\textwidth}{!}{
\begin{tabular}{cccccccc}
\toprule
\textbf{Subject} & \textbf{sLDA (0,6s)} & \textbf{sLDA (0,8s)} & \textbf{sLDA (1s)} & \textbf{RF (0,6s)} & \textbf{RF (0,8s)} & \textbf{RF (1s)} & \textbf{CNN}  \\
\midrule
G01              & 0,65              & 0,65              & 0,65            & 0,61            & 0,59            & 0,67          & \textbf{0,79} \\
G02              & 0,61              & 0,55              & 0,53            & 0,49            & \textbf{0,63}   & 0,47          & 0,43          \\
G03              & 0,51              & 0,59              & \textbf{0,67}   & 0,59            & 0,53            & 0,53          & 0,58          \\
G04              & \textbf{0,67}     & 0,63              & 0,61            & 0,53            & 0,59            & 0,55          & 0,58          \\
G05              & 0,52              & 0,63              & 0,65            & 0,58            & 0,63            & 0,56          & \textbf{0,75} \\
G06              & 0,51              & 0,60              & 0,64            & \textbf{0,68}   & 0,53            & 0,49          & 0,55          \\
G07              & 0,58              & 0,50              & 0,54            & 0,58            & \textbf{0,60}   & 0,50          & \textbf{0,60} \\
G08              & \textbf{0,78}     & 0,75              & 0,78            & 0,67            & 0,76            & 0,73          & 0,72          \\
G09              & 0,69              & 0,69              & 0,62            & 0,58            & 0,71            & 0,63          & \textbf{0,73} \\
G10              & \textbf{0,65}     & 0,54              & 0,54            & 0,63            & 0,61            & 0,63          & \textbf{0,65} \\
G11              & 0,56              & 0,60              & \textbf{0,63}   & 0,58            & 0,50            & 0,56          & 0,61          \\
G12              & 0,56              & 0,58              & 0,58            & 0,66            & 0,66            & 0,52          & \textbf{0,80} \\
G13              & 0,53              & 0,45              & 0,55            & 0,51            & \textbf{0,63}   & 0,51          & 0,57          \\
G14              & 0,64              & 0,68              & \textbf{0,75}   & 0,55            & 0,59            & 0,55          & 0,60          \\
G15              & 0,63              & \textbf{0,71}     & 0,59            & 0,57            & 0,61            & 0,55          & 0,65          \\
\midrule
\textbf{MEAN}    & 0,61              & 0,61              & 0,62   & 0,59            & 0,61            & 0,56          & \textbf{0,64}         \\
\textbf{STD}    & 0,08              & 0,08              & 0,07             & 0,06            & 0,07            & 0,07          & 0,10         \\
\bottomrule
\end{tabular}}
\label{tab:AccComp2}
\end{table*}

\subsection{Classification results}
Tab. \ref{tab:AccComp1} and Tab. \ref{tab:AccComp2} report the comparison of the classification performance between the \gls{cnn} and the baseline models over the unseen validation set of the two datasets.
They show the results of the classification in terms of accuracy. To achieve these performance, we used the \gls{cnn} model with the best selection of hyperparameters, employing the same architecture for all participants. On the other hand, for \gls{slda} and \gls{rf} we considered all possible choice of the sliding window lenght, with the best window time location, for each participant.
The chance level was computed as in Section \ref{sec:eval} and it was found to be $0.40$.
Comparing the classification results among the three classifiers, we can see similar accuracies for all of them, with all values above the chance level.
We can also notice that they achieved slightly better results in the \emph{Dataset 1}, as expected from its higher repeatibility across \gls{eeg} segments (see Fig.\ref{fig:mrcp}), compared to the \emph{Dataset 2}.
However, for both datasets, the \gls{cnn} model reached the best average accuracy across all participants ($0.70$ for \emph{Dataset 1}, $0.64$ for \emph{Dataset 2}).
\gls{slda} and \gls{rf} achieved the best classification accuracy, at the single-subject level in \emph{Dataset 1}: thus, a particular configuration (i.e., an optimal choice of the window length and time location) can lead a baseline model to yield higher performance compared to \gls{cnn}.
%
%
Nevertheless, especially for the \textit{Dataset 2}, the \gls{cnn} showed higher variability in the individual participant accuracies, with some of them reaching very high values ($0.80$ for G12) and others slightly above the chance level ($0.43$ for G02).
Finally, from the confusion matrices (not reported here for space constraints), we observed that the rest class was classified with the highest accuracy compared to the other movement classes (best accuracy among the two datasets: 78\% for rest, 57\% for touch, 62\% for grasp, 55\% for palmar, 52\% for lateral), in line with previous literature \cite{ofner2017upper, schwarz2017decoding, schwarz2019unimanual}.
As expected, the computational complexity for \gls{slda} and \gls{rf} is significantly lower compared to \gls{cnn}: the former use less time points as input to train the model (either $0.6$s, $0.8$s and $1$s) and a shorter training time, while the latter took the entire $5$s \gls{eeg} segments into account and a longer time to train.
However, \gls{cnn} showed promising advantages over \gls{slda} and \gls{rf}: indeed, to reach comparable performance, the latter exploited a semi-quantitative pre-processing pipeline, including \gls{ica} to clean data from eye movements artifacts. Moreover, they had to train a classifier at each time point to select the one that led to the best performance. On the other hand, lighter pre-processing is needed to classify the datasets by means of the \gls{cnn}; and it is completely automatic.
Even if two relatively small datasets were available, we could show that our \gls{cnn} model can achieve classification accuracies in line with two well-established baseline models.
Moreover, we obtain similar performance with a simpler pre-processing pipeline, reducing it to those steps (e.g., filtering and automatic trial rejection) that could be performed in an online modality.
This may be explained by the fact that the \gls{cnn} can both behave as an automatic feature extraction method, and as an efficient classifier.
Finally, \gls{cnn} could take larger advantage by the spatial information in the \gls{eeg} dataset, by applying a spatial convolution at its second layer. On the other hand, \gls{slda} and \gls{rf} did not use this kind of information to enhance their predictions.

%
%
%
%


 \section{Conclusions}\label{sec:conclusions}
 In this study, we considered two different datasets where fine hand movements (touch, grasp, palmar and lateral grasp) were repeated in a self-paced modality, and we evaluated the classification performance of a deep learning model, i.e., a \gls{cnn}, into respect to two well-established machine learning models, i.e., \gls{slda} and \gls{rf}.
 The classification included three classes, i.e., two movements and the rest condition, and it was based on the components of the \gls{eeg} signals in the $0.3$-$3$ Hz low frequency band. This is the typical band to detect the \gls{mrcp}.
 We showed that \gls{cnn} achieved good performance in both datasets (average accuracy of $0.70$ in \emph{dataset 1}, $0.64$ in \emph{dataset 2}, with a chance level of $0.40$), and they were similar or superior to the baseline models.
 All classifiers yielded better results in the first dataset (touch, grasp and rest), in line with the electrophysiological observations on the \gls{mrcp} that were more pronounced in that dataset. Moreover, for similar reasons, the rest condition always led to the highest true positive rate.
 We also highlighted that our \gls{cnn} did not require the use of \gls{ica}, that is a common, but heavy burden and semi-quantitative pre-processing step, paving the way for its possible use in an online modality, e.g., in many \gls{bci} applications.

 \section*{Acknowledgments}
This work was partly supported by EU Horizon 2020 Project MoreGrasp ('643955').
Part of this work was also supported by MUR (Italian Minister for University and Research) under the initiative "Departments of Excellence" (Law 232/2016).
 The authors thank Andreas Schwarz for his effort designing, implementing and adjusting the paradigm, as well as Sophie Zentner for data recording.

\bibliography{biblio}
\bibliographystyle{IEEEtran}

\end{document}